\pdfoutput=1	

\documentclass{elsart3p}
\usepackage{graphicx}
\usepackage{amssymb}
\usepackage{ifthen}

\begin{document}


\newcommand{\rmi}{\mathrm{i}}
\newcommand{\rmd}{\mathrm{d}}
\newcommand{\bi}[1]{\mathbf{#1}}

\newcommand{\METATOY}{\underline{meta}ma\underline{t}erial f\underline{o}r ra\underline{y}s}

\ifthenelse{
	\equal{false}	
	{true}
}
{
	\newcommand{\change}[2]{\textbf{#1 [#2]}}
	\newcommand{\addition}[1]{\textbf{#1}}
	\newcommand{\comment}[1]{\textbf{[#1]}}
}
{
	\newcommand{\change}[2]{#1}
	\newcommand{\addition}[1]{#1}
	\newcommand{\comment}[1]{}
}

\begin{frontmatter}

\title{Experimental demonstration of a light-ray-direction-flipping METATOY based on confocal lenticular arrays}

\author{Michael Blair, }
\author{Leo Clark, }
\author{E.\ Alasdair Houston, }
\author{Gary Smith, }
\author{Jonathan Leach, }
\author{Alasdair C.\ Hamilton, and}
\author{Johannes Courtial}
\ead{j.courtial@physics.gla.ac.uk}
\address{Department of Physics and Astronomy, Faculty of Physical Sciences, University of Glasgow, Glasgow G12~8QQ, UK}

\begin{abstract}
We show, theoretically and experimentally, that a sheet formed by two confocal lenticular arrays can flip one component of the local light-ray direction.
Ray-optically, such a sheet is equivalent to a Dove-prism sheet, an example of a METATOY (\change{\METATOY}{metamaterial for light rays}), a structure that changes the direction of transmitted light rays in a way that cannot be performed perfectly wave-optically.
\end{abstract}



\end{frontmatter}

\section{Introduction}

A metamaterial is often defined as a structure with electromagnetic properties not usually observed in nature \cite{Sihvola-2007}.
Built to interact with electromagnetic waves, metamaterials are (usually periodic) structures whose unit cells -- often called ``meta-atoms'' -- are on a scale smaller than the wavelength of the incident electromagnetic waves.
Metamaterials can, for example, have negative permittivity, $\epsilon$, and permeability, $\mu$, \cite{Smith-et-al-2000} giving them a negative refractive index \cite{Veselago-1968}.
Such materials were first demonstrated for microwaves \cite{Shelby-et-al-2001}; efforts to realize them for visible wavelengths are ongoing \cite{Boltasseva-Shalaev-2008}.


Optical components on a scale much greater than the optical wavelength, for example prisms, may also be joined together in a periodic structure \cite{Hamilton-Courtial-2009}.
These so-called METATOYs \cite{Wikipedia-METATOY} \change{can partially}{are designed to} mimic the ray optics of metamaterials\addition{, as expressed in the acronym METATOY, which stands for \METATOY.}
\change{An example is}{, for example} the bending of light rays like the interface between optical media with refractive indices of opposite signs \cite{Courtial-Nelson-2008}\change{.}{;}
In any case, \change{METATOYs}{they} are structures with ray-optical properties not usually observed in nature.
Although they fit some definitions of metamaterials, METATOYs are not metamaterials in the usual sense, and they do not have wave-optical-metamaterial properties such as amplification of evanescent waves that can turn a sliver of negative-refractive-index metamaterials into a perfect lens~\cite{Pendry-2000}.

\comment{moved from middle of previous paragraph}
\change{It}{it} can even be argued that \change{METATOYs can}{they} perform light-ray-direction changes that create light-ray fields without wave-optical analog in the ray-optics limit \cite{Hamilton-Courtial-2009}.
\addition{This property of METATOYs has the potential to open up new possibilities in fields that are traditionally limited by wave optics.
One such field is optical imaging, which is traditionally limited by wave optics in the sense that most theorems are derived from the wave-optical principle of equal optical path (see, for example, Ref.\ \cite{Born-Wolf-1980-principle-of-equal-optical-path}).  There are already specific examples of novel imaging with METATOYs \cite{Courtial-2008a,Hamilton-Courtial-2008c}; Ref.\ \cite{Courtial-2009b} is the start of a systematic exploration of optical imaging not limited by wave optics.}

One of the simplest examples of a METATOY is a Dove-prism array \cite{Hamilton-Courtial-2008a}.
A Dove-prism array flips one component of the local light-ray direction \cite{Hamilton-Courtial-2008a}, which is why we call it here a ray-flipping sheet.
Ray-flipping sheets form the basis of a ray-rotation sheet \cite{Hamilton-et-al-2009}, a structure that usually serves as our favourite example of a METATOY.
A ray-rotation sheet was, for example, used to demonstrate the ability of METATOYs to produce light-ray fields without wave-optical analog in the ray-optics limit \cite{Hamilton-Courtial-2009}; forced application of the wave-optically motivated Fermat's principle to ray-rotation sheets leads to a formal equivalence with the interface between two optical media with a complex refractive-index ratio \cite{Sundar-et-al-2009}.

Another example of a METATOY consists of confocal lenslet arrays (that is, confocal arrays of spherical microlenses) \cite{Courtial-2008a}.
In the simplest case, namely if the two lenslet arrays have the same focal length and are aligned such that corresponding lenslets in the two arrays share the same optical axis, such confocal lenslet arrays act -- ray-optically -- like the interface between two optical media with refractive indices of equal magnitude but opposite sign, and they perform pseudoscopic imaging (that is, imaging in which depth appears inverted).
This has been realized experimentally \cite{Stevens-Harvey-2002,Okano-Arai-2002}.
Different cases of two confocal lenslet arrays with the same focal length but aligned such that corresponding lenslets in the two arrays do not share the same optical axis, or in which the alignment varies across the arrays, have also been realized experimentally.
These include the moir\'{e} magnifier \cite{Hutley-et-al-1994}, which comprises confocal spherical-lenslet arrays that are rotated with respect to each other, and microoptical imaging systems \cite{Volkel-et-al-2003,Duparre-et-al-2005}, which comprise confocal spherical-lenslet arrays with different pitches (in addition to an array of field lenses).

Here we show that a ray-flipping sheet can alternatively be built from a pair of confocal lenticular arrays (that is, confocal cylindrical lenslet arrays).
We realize a ray-flipping sheet in the form of confocal lenticular arrays experimentally and present photos of the view through it, confirming a few of the simulations from Ref.\ \cite{Hamilton-Courtial-2009}.

\section{How confocal lenticular arrays flip one transverse light-ray direction\label{Comparison}}

\begin{figure}
\centering
\includegraphics{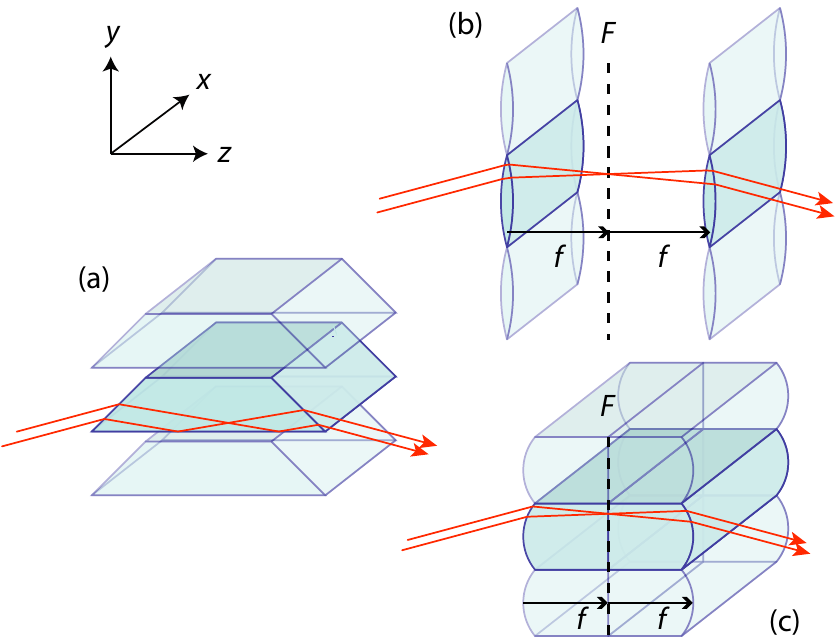}
\caption{\label{Dove-prisms-and-cyl-lenses-figure}(a) A Dove-prism array -- Dove prisms stretched in the $x$ direction and stacked in the $y$ direction -- flips the direction of transmitted light rays in the $y$ direction.
(b)~Two arrays of cylindrical lenses that share a common focal plane, $F$, also flip the $y$ component of the direction of transmitted light rays.  The focal length of each cylindrical lens is $f$.
(c)~In our experiment, we use back-to-back commercial lenticular arrays.
The grooved front surface of each lenticular array acts as a cylindrical-lens array whose focal plane, $F$, coincides with the planar back surface.}
\end{figure}

Light passing from left to right through a Dove prism aligned as in Fig.\ \ref{Dove-prisms-and-cyl-lenses-figure}(a) is refracted at the left face, reflected (through total internal reflection) at the bottom face, and then again refracted at the right face.
In the coordinate system of Fig.\ \ref{Dove-prisms-and-cyl-lenses-figure}, transmission of a light ray through a Dove prism then has the following effects on the light ray's position and direction.
The position at which the light ray exits the Dove prism can be offset with respect to the entry position in the $x$ and $y$ directions.
Due to the prism's symmetry, the angle of refraction at the left face and the angle of incidence at the right face have the same magnitude and opposite sign, and so the angle of refraction at the right face has the same magnitude as the angle of incidence at the entrance face, but the opposite sign.
Overall, the direction of a transmitted light ray remains the same, apart from an inversion of the $y$ component of the direction of a transmitted light ray.

Stacked Dove prisms form a Dove-prism array \cite{Lian-Chang-1996}.
Fig.\ \ref{Dove-prisms-and-cyl-lenses-figure}(a) shows a Dove-prism array in which the Dove prisms are stacked in the $y$ direction.
Compared to a standard Dove prism, which has a square cross section, the Dove prisms shown in Fig.\ \ref{Dove-prisms-and-cyl-lenses-figure}(a) are stretched in the $x$ direction; the Dove-prism array then forms a sheet parallel to the $(x,y)$ plane.
If the Dove prisms are shrunk, the maximum $(x,y)$ offset introduced on transmission is shrunk by the same factor (as it is limited by the aperture size of the individual prisms) while the effect on the light-ray direction remains unchanged.
For some applications, e.g.\ visual applications, the prisms can be shrunk sufficiently for the offset to become negligible.
The sole effect on transmitted light rays of such a Dove-prism array is then an inversion of the $y$ component of the directions of transmitted light rays \cite{Hamilton-Courtial-2008a}.

Building a Dove-prism array in practice is either cumbersome, relatively expensive, or both.
An easier alternative is to use two arrays of cylindrical lenses -- lenticular arrays -- that are separated by the sum of their focal lengths (they are confocal; see Fig.\ \ref{Dove-prisms-and-cyl-lenses-figure}(b)).
Lenticular arrays are inexpensive as they are frequently used for lenticular printing, a technique that can, for example, produce an illusion of depth in a flat image \cite{Wikipedia-lenticular-lens}.
Furthermore, they are manufactured as a single sheet of plastic with a corrugated side (the ridges form the cylindrical lenses) and a flat side that coincides with the cylindrical lenses' focal plane, so it is relatively straightforward to arrange two lenticular arrays such that they share a common focal plane (Fig.\ \ref{Dove-prisms-and-cyl-lenses-figure}(c)).

Transmission through a pair of confocal cylindrical lenses aligned as shown in Fig.\ \ref{Dove-prisms-and-cyl-lenses-figure}(b) inverts the $y$ component of the light-ray direction, while not affecting the $x$ component.
If the cylindrical lenses are parts of two lenticular arrays in which the individual cylindrical lenses are sufficiently small to make the offset also introduced by transmission through the lenses negligible, then the effect on light rays transmitted through corresponding cylindrical lenses is the same as that of transmission through a Dove-prism array, namely inversion of one transverse component of the light-ray direction.

\section{Experiment}


\begin{figure}[ht]
\centering
\includegraphics{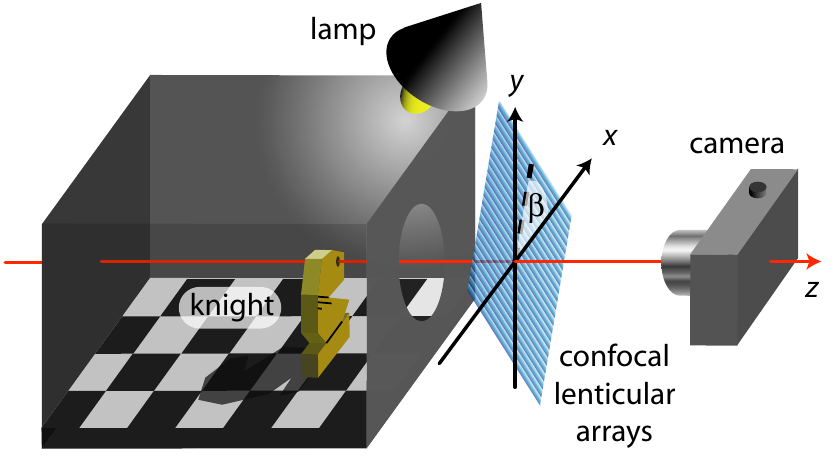}
\caption{\label{setup-figure}Schematic of the experimental setup.
A chess piece (Fig.\ \ref{knights-figure}) was placed a variable distance (``object distance'') behind confocal lenticular arrays and was then photographed with a camera, placed a fixed distance (``camera distance'', 96.3cm) in front of the confocal lenticular arrays.
A HeNe laser beam (switched off during photo taking) defined the $z$ axis.
The chess piece was centered on the $z$ axis as shown; the camera lens was centered on the $z$ axis.
The confocal lenticular arrays were placed perpendicular to the $z$ axis, defining the $(x,y)$ plane.
The arrays could be rotated around the $z$ axis, so that the flip direction could have an arbitrary angle, $\beta$, with the $x$ axis.}
\end{figure}

\begin{figure}
\centering \includegraphics{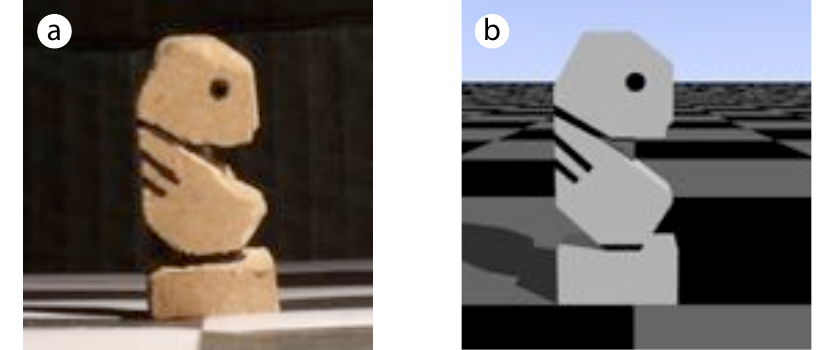}
\caption{\label{knights-figure}Home-made chess piece (a) and its imperfect representation in POV-Ray~(b).
In Figs \ref{fig:KnightRotate} and \ref{fig:KnightMove}, this chess piece served as the object that was viewed through ray-flipping sheets.}
\end{figure}

A diagram of our experimental setup is shown in Fig.\ \ref{setup-figure}.
Two lenticular arrays (Edmund Optics B43-028, $f=0.085$in) were held in a rotatable rigid mount (not shown) with their flat faces touching. 
The mount allowed alignment of the two arrays relative to each other; alignment was judged to be optimal when a laser beam at normal incidence passed through the two arrays without deviation.
The lenticular arrays then formed a useful approximation to a ray-flipping sheet.
A camera (Canon EOS 450D with Canon EF 100mm F/2.8 macro lens) was positioned to one side of the lenticular arrays pointing normally to its plane and at the centre of it.
A home-made chess piece (Fig.\ \ref{knights-figure}), placed at the other side of the array, served as the main object to be viewed through the lenticular arrays.

\begin{figure}
\centering
\includegraphics{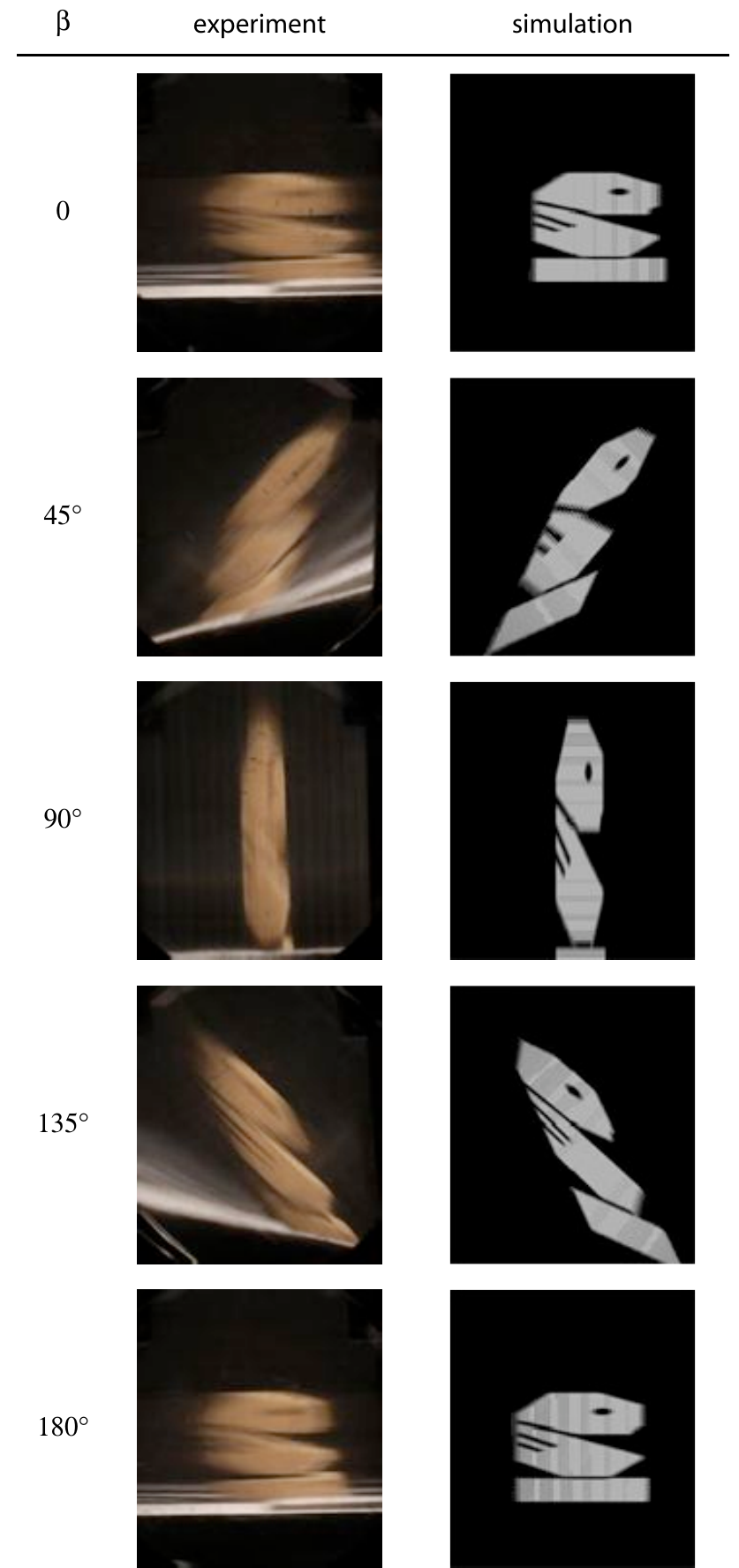}
\caption{\label{fig:KnightRotate}View through a ray-flipping sheet for various orientations of the sheet.
Different rows correspond to different sheet orientations, described by the angle $\beta$ between the flip direction and the horizontal (Fig.\ \ref{setup-figure}).
From left to right, the columns show: the value of $\beta$, photos of a scene including a chess piece taken through confocal lenticular arrays, and the simulated view through a corresponding Dove-prism sheet.
The experiment was performed for a camera distance of 96.3cm (measured to the detector plane) and an object (chess piece) distance of 43.7cm.
In the simulation, all objects other than the chess piece were removed for clarity.
The simulation was performed using the ray-tracing software POV-Ray \cite{POV-Ray}.}
\end{figure}

\begin{figure}
\centering
\includegraphics{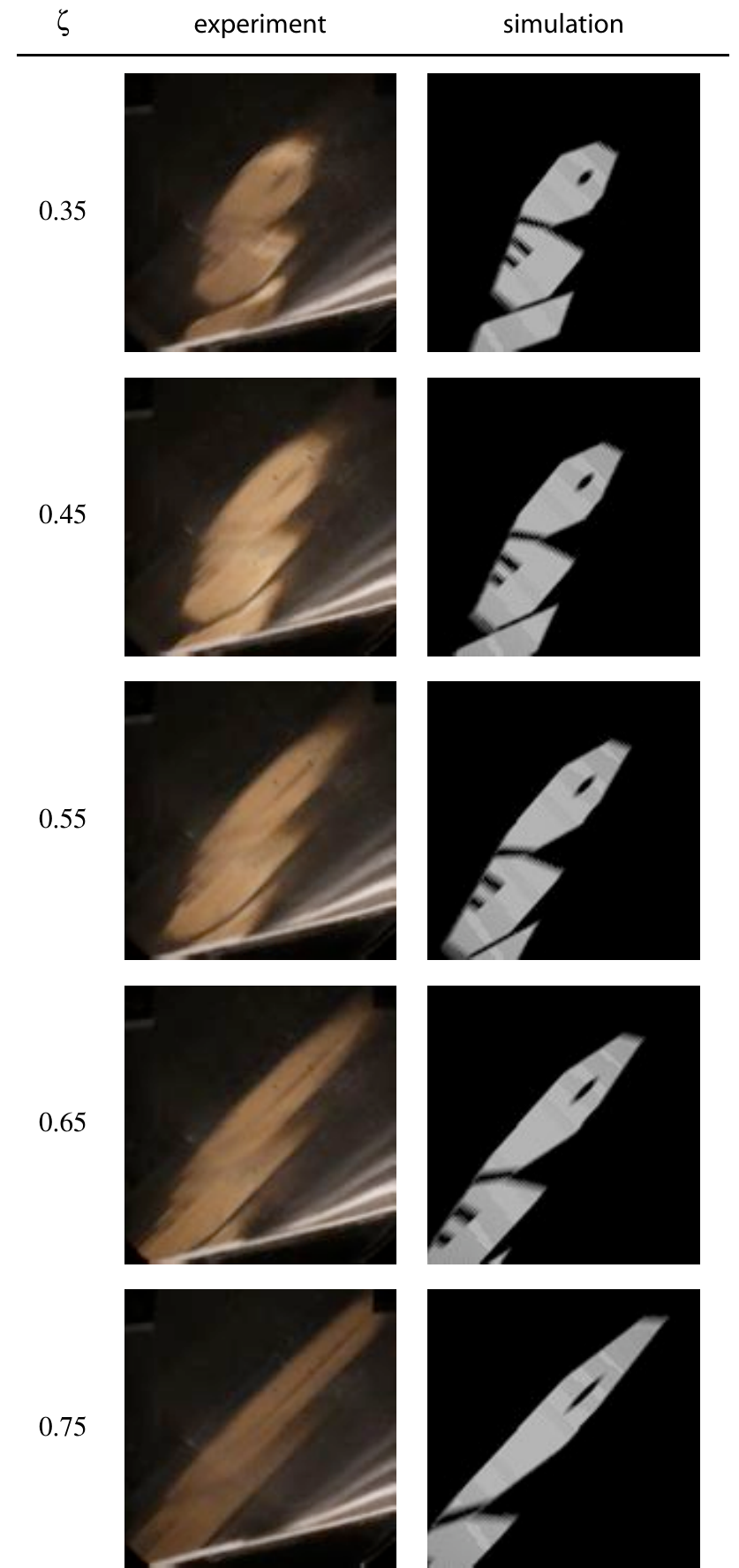}
\caption{\label{fig:KnightMove}View through a ray-flipping sheet for various ratios $\zeta$ between object distance and camera distance.
Different rows correspond to different values of $\zeta$.
The value of $\zeta$ is shown in the left column.
The two picture columns show experimental photos (center) and ray-tracing simulations (right).
The angle between the ray-flipping sheet and the horizontal was $\beta = 45^\circ$, the camera distance was 96.3cm.}
\end{figure}
  
Figures \ref{fig:KnightRotate} and \ref{fig:KnightMove} show photos of our chosen object, taken through our experimental realization of a ray-flipping sheet, and compare these with ray-tracing simulations (performed using the freeware program POV-Ray \cite{POV-Ray}) through the detailed structure of a ray-flipping sheet in the form of a Dove-prism array.
Following Ref.\ \cite{Hamilton-Courtial-2008a}, one parameter was altered while the other parameters were kept constant.
In Fig.\ \ref{fig:KnightRotate}, the sheet -- and with it the flip direction -- was rotated around the $z$ axis while keeping the object distance fixed.
As expected for this particular ratio between object distance and camera distance, the object appears stretched in the direction of the flip direction.
In Fig.\ \ref{fig:KnightMove}, the object distance was changed, keeping the sheet orientation fixed.
For practical reasons, the object distance was not increased above the camera distance.
As the object distance is increased, the factor by which the object appears stretched in the flip direction is also increased.
In both figures, the view through the confocal lenticular arrays appears more blurred than it does in the corresponding simulations.
We believe this is due to imperfections in the imaging quality of the lenticular arrays and in the alignment of the arrays relative to each other.
\addition{(Our confocal lenticular arrays also suffer from field-of-view limitations which are discussed theoretically in some detail in Ref.\ \cite{Courtial-2009}.  We have designed our experiment such that its field of view is not limited in this way; specifically, we have chosen the object distance and camera distance to be significantly greater than the object size (the height of the chess piece is approximately 4.7cm).)}
Nevertheless, in both figures there is good overall agreement between experiment and simulations.

\begin{figure}
\centering \includegraphics{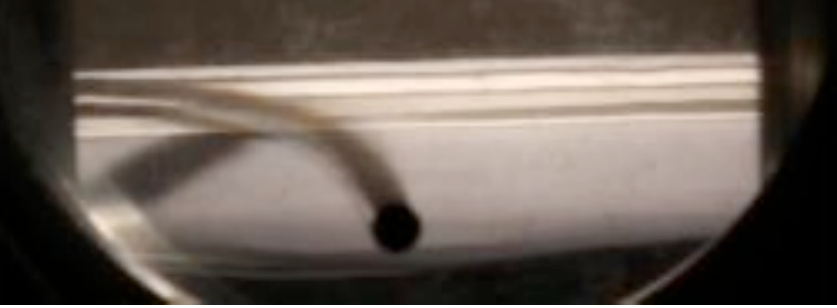}
\caption{\label{hyperbola-figure}Experimental test of the prediction straight line perpendicular to a ray-flipping sheet, when viewed through the sheet, appears bent into a hyperbola \cite{Hamilton-Courtial-2008a}.
The photo shows a straight metal bar that is approximately perpendicular to the ray-flipping sheet, seen through confocal lenticular arrays orientated such that they invert the horizontal light-ray-direction component ($\beta = 0^\circ$).
The end of the bar (small back disc) touches the ray-flipping sheet.}
\end{figure}

Figure \ref{hyperbola-figure} shows the result of a quick check of the prediction that a straight line perpendicular to a ray-flipping sheet appears bent into a hyperbola when seen through the sheet \cite{Hamilton-Courtial-2008a}.
A straight metal bar was used as an object approximating a straight line.
Due to imperfections in the experiment which we intend to investigate and eliminate in future experiments, details of the apparent bending of the metal bar are not as predicted; specifically, the point where the bar touches the ray-flipping sheet appears slightly too low on the hyperbola.
Nevertheless, the metal bar does appear bent into a hyperbola, as predicted.

\section{Conclusions and future work}

It is reassuring to see the predictions of calculations and computer simulations confirmed experimentally, and this paper provides the first experimental confirmation of the visual properties of ray-flipping METATOYs.
Our experiments show up imperfections in the specific implementation we have chosen, but we believe that we can significantly improve on these first experiments.

Most importantly, we see these ray-flipping sheets as an important step towards the realization of significantly more complex METATOYs.
Specifically, we are currently working on experimental realizations of ray-rotation sheets \cite{Hamilton-et-al-2009}, confocal lenslet arrays that mimic a refractive-index interface \cite{Courtial-2008a}, and of generalized confocal lenslet arrays \cite{Hamilton-Courtial-2009b}.

\section*{Acknowledgments}

ACH is supported by the UK's Engineering and Physical Sciences Research Council (EPSRC).
JC is a Royal Society University Research Fellow.


\bibliographystyle{osajnl}
\bibliography{/Users/johannes/Documents/work/library/Johannes}

\end{document}